\newcommand{\beq}{\begin{equation}}
\newcommand{\eeq}{\end{equation}}

\newcommand{\bear}{\begin{eqnarray}}
\newcommand{\eear}{\end{eqnarray}}

\documentclass[a4paper]{jpconf}
\usepackage{graphicx}
\usepackage{amssymb}
\begin{document}
\title{A new amplification structure for time projection chambers based on electroluminescence}

\author{D. Gonz\'alez-D\'iaz, M. Fonta\'i\~ na, D. Garc\'ia Castro}

\address{Instituto Galego de F\'isica de Altas Enerx\'ias, Univ. de Santiago de Compostela, Campus Vida, R\'ua Xos\'e Mar\'ia Su\'arez N\'u\~nez, s/n, Santiago de Compostela, E-15782, Spain}

\author{B. Mehl, R. de Oliveira, S. Williams}
\address{CERN, Esplanade des Particules 1, Meyrin, Switzerland}

\author{F. Monrabal}
\address{Donostia International Physics Center (DIPC), Paseo Manuel Lardizabal, 4, Donostia-San Sebastian, E-20018, Spain}

\author{M. Querol, V. \'Alvarez}
\address{Instituto de F\'isica Corpuscular (IFIC), CSIC \& Universitat de Valencia, Calle Catedrá\'atico Jos\'e Beltr\'an, 2, Paterna, E-46980, Spain}

\ead{Diego.Gonzalez.Diaz@usc.es}

\begin{abstract}
A simple hole-type secondary scintillation structure (2~mm-hole, 5~mm-pitch, 5~mm-thickness) is introduced and its operation demonstrated in pure xenon in the pressure range 2-10~bar. The new device, characteristically translucent, has been manufactured through a collaboration between IGFAE and the CERN workshop, and relies entirely on radiopure materials (acrylic and copper), being extremely rugged in the presence of sparks, mechanically robust, and easily scalable, yet made through a relatively simple process. With an overall figure (at 10~bar) characterized by an energy resolution of 18.9\%(FWHM) for $^{55}$Fe x-rays, an optical gain of m$_\gamma$ = 500~ph/e, and a stable operation at reduced fields more than twice those of some of the presently running experiments ($E_{EL}=3$~kV/cm/bar), this family of structures seems to show great promise for electroluminescence readouts on large scale detectors. As argued below, further improvements have the potential of bringing the energy resolution close to the Fano factor and increasing the optical gain.
\end{abstract}

\section{Introduction}

Particle detectors based on secondary scintillation such as optical time projection chambers (OTPCs) or electroluminescence ones (ELTPCs), usually developed for the study of rare interactions, are developing at considerable speed over the last 10-15 years. They currently lead the direct searches of WIMP dark matter (XENON \cite{XENON}, LZ \cite{LZ}, PandaX \cite{PANDAX} and DarkSide \cite{DarkSide}), representing one of the most promising techniques for achieving directional information (DMTPC \cite{DMTPC}, CYGNO \cite{CYGNO}); in nuclear physics they are employed in the study of $\beta\beta0\nu$ and $\beta\beta2\nu$ decays (NEXT \cite{NEW}), photo-production of high-lying Hoyle states (TUNL-TPC \cite{TUNL}), 2p-decay and related processes (Warsaw-TPC \cite{Warsaw}); very recently, they have been proposed for the study of neutrino oscillations (ARIADNE \cite{ARIADNE}). This growth has not been paralleled, however, by a comparable development of the associated scintillation structures, largely consisting of meshes or, more rarely, conventional micro-pattern gas detectors (MPGDs) optimized for operation in avalanche mode under quenched gases. The lack of dedicated amplification structures is consequential in view of the presumption of scalability for most of the aforementioned detectors, that is at odds with the increased practical problems encountered with the handling, alignment, and stretching of many-m$^2$ meshes or wires. Furthermore, when used in conjunction with lenses, the number of optical sensors needed is inversely proportional to the achievable optical gain, if not limited by space resolution (see, e.g., \cite{DiegoRev}). An optimized structure can allow higher optical gains in conditions where no avalanche multiplication takes place, thus eliminating ion back-flow and improving the energy resolution and detector stability.

We propose a purposeful alternative to the use of meshes or thick-GEMs, based on an acrylic (Poly(methyl methacrylate), PMMA) plate with two thermally bonded electrodes, on which a hatched pattern is made by photo-lithography, to ensure a degree of translucence. For the typical thicknesses required in EL-chambers ($\gtrsim5$mm-scale), high electron transmission through mm-diameter holes, at 5~mm pitch, can be anticipated. Given the sturdiness of acrylic at this thickness, there is no need of machining glass fiber epoxy plates as in the case of thick-GEMs, anticipating a better surface finish. Unlike those, and glass or ceramic GEMs, acrylic GEMs are based on radiopure materials and, compared to CuFlon GEMs, they are transparent and affordable in large areas. The operating temperature range of acrylic is at least 77-393~K. For applications based on electroluminescence, there is indeed no apparent performance trade-off for the above advantages, except in the presence of potentially strong outgassing in the holes or surface discharges, that we show here not to be the case. In a way, it seems almost striking that such a simple device has not been built before.

\begin{figure}[h!!!]
\begin{center}
\includegraphics[width=12.5cm]{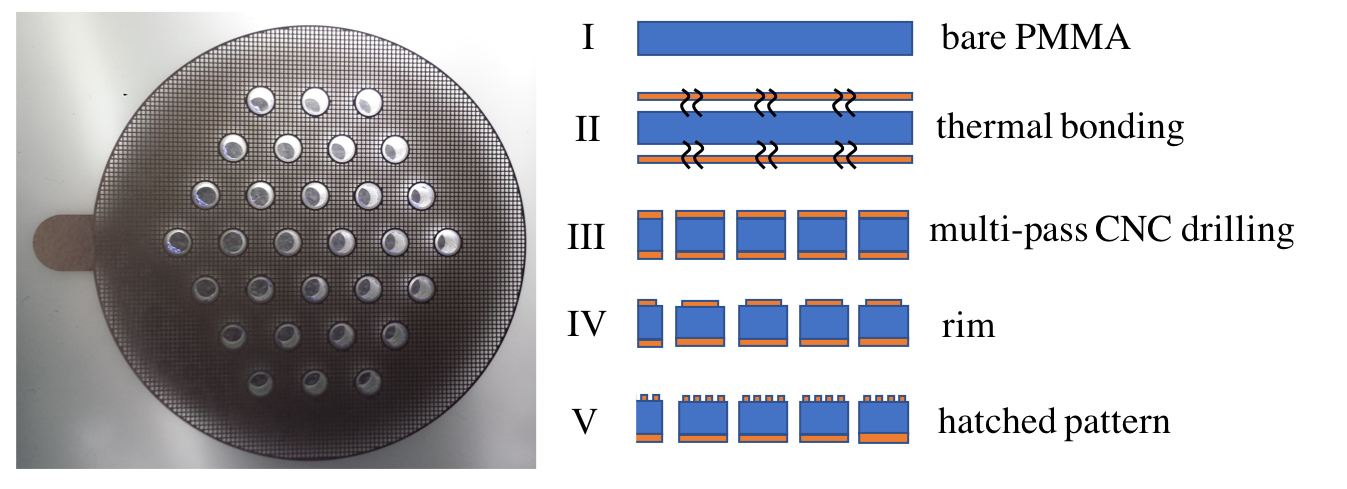}
\end{center}
\caption{\label{photo_FAT} Left: photograph of the 2~mm-hole/5~mm-pitch FAT-GEM characterized in this work. The structure is illuminated from below, that allows to infer its translucent nature to photons produced inside the holes. Right: description of the fabrication process, followed at the RD51 workshop at CERN. Step IV is done in a chemical bath and step V by photo-lithography. In the photograph, some leftovers of the photoresist can be noticed inside the holes, however the process will be optimized in the next production batch.}
\end{figure}

We christened the structure as FAT-GEM (Field-Assisted Transparent Gaseous Electroluminescence Multiplier),\footnote{The structure is not \emph{inherently} transparent to VUV light, but it is so for CF$_4$ scintillation. Besides, we target at a high device transparency by means of physical evaporation of tetraphenyl butadiene (TPB) inside the channels.} and present here the first competitive results, with a more systematic study of the behaviour for different geometries and operating conditions to be reported in the near future.

\section{Description of the amplification structure}

\begin{figure}[h!!!]
\begin{center}
\includegraphics[width=14cm]{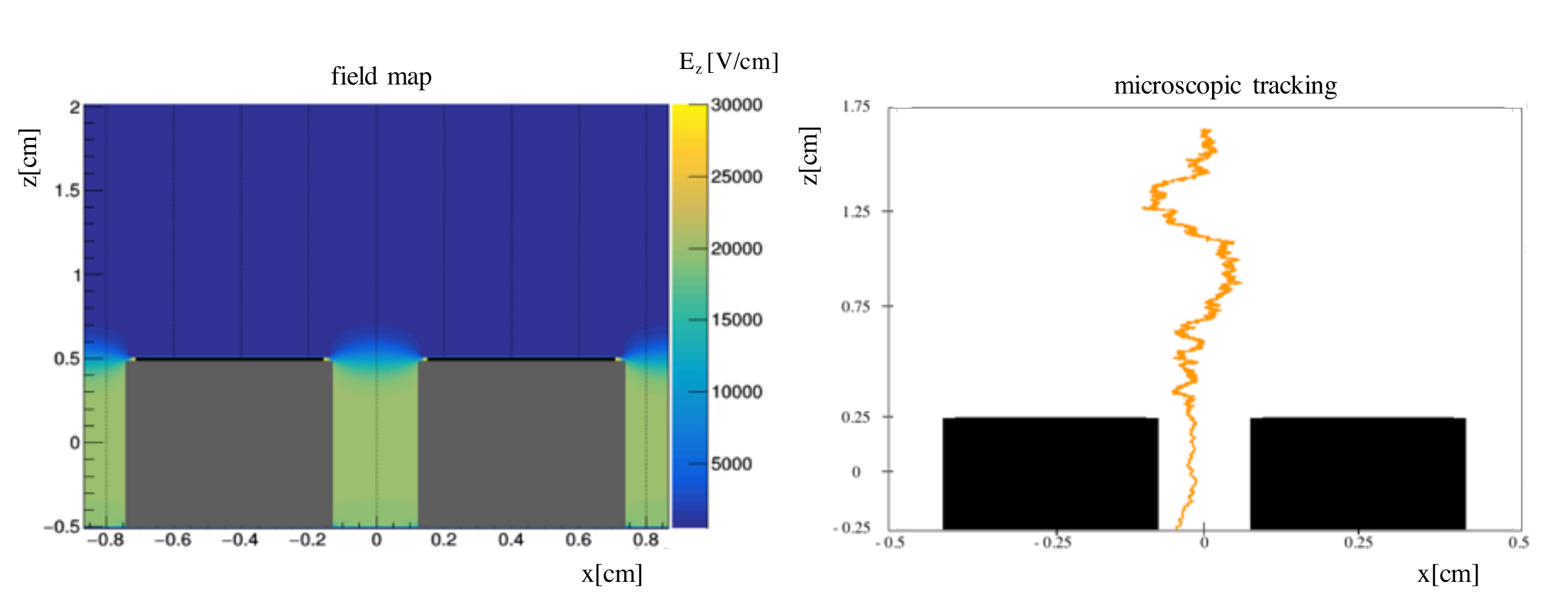}
\includegraphics[width=14cm]{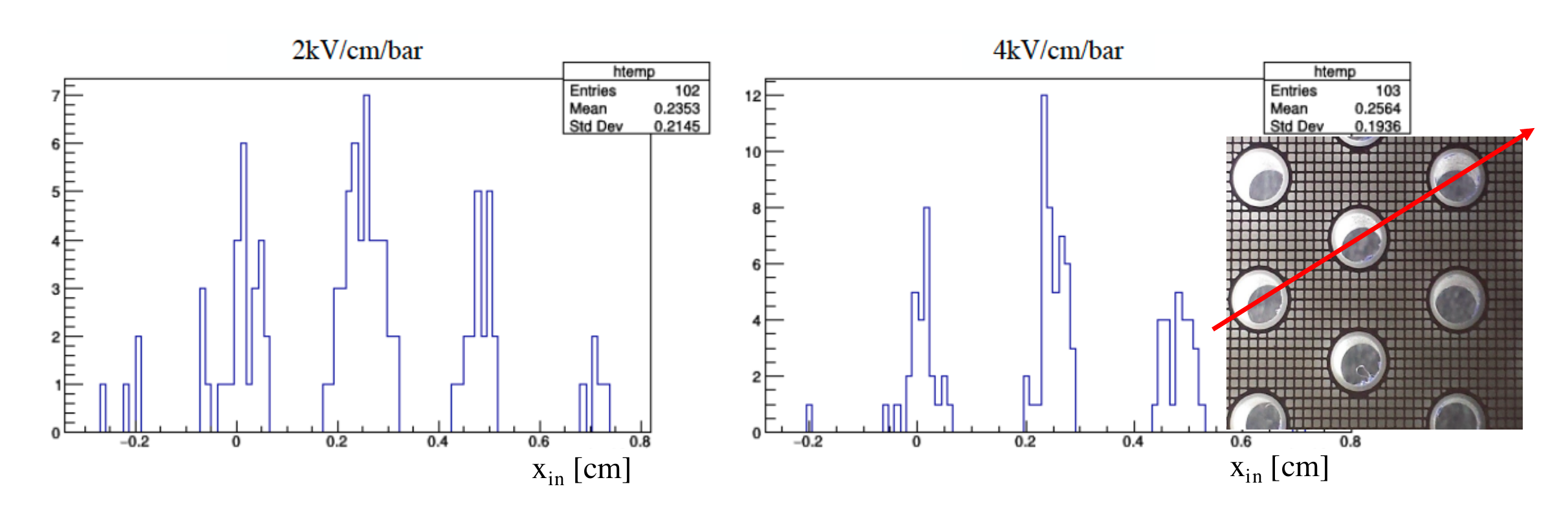}
\end{center}
\caption{\label{FirstSim} Top-left: map of the longitudinal component of the electric field at the holes mid-plane, obtained with Gmsh and ELMER \cite{Gmsh, ELMER}. Top-right: reconstructed trajectory from the position of the elastic scatters obtained with Garfield++, by resorting to Gmsh+ELMER maps. Bottom: illustration of the field-focusing effect, by plotting the $x$ coordinate at the hole entrance, along a direction like the one indicated in the photograph.}
\end{figure}

The present fabrication procedure of the FAT-GEMs is sketched in Fig.\ref{photo_FAT}-right, and a photograph of the structure characterized in this work is shown in left. It has been designed, as customary, with the help of the Garfield++ simulation code \cite{Gar}, enhanced through new packages for the simulation of scintillation in pure gases \cite{Oliveira} and its mixtures \cite{MicroSim}. The baseline design and requirements are obtained from the parameters of the NEW detector (first phase of the NEXT $\beta\beta0\nu$ experiment, \cite{NEW}), i.e., a 5~mm scintillation gap, a hole pitch adequate for few mm-scale spatial sampling (i.e., compatible with operation under low-diffusion gases \cite{Henr}), a drift field around $E_d = 50$~V/cm/bar, and a pressure of 10~bar of xenon. For convenience, the active area was designed to be approximately 35~mm in diameter, with the metalization extending up to 55~mm. We used a drift region of 15~mm, and a short-range x-ray source ($^{55}$Fe) placed at the cathode, so that most events come from within few mm's of the cathode region at most, where field lines are uniform (Fig. \ref{FirstSim} top-left).

Fig. 2-top conveys the longitudinal field map (left) and a representation of one electron track (right) for two different geometries. Simulations have been done with a continuous anode (implemented through a 71\% transparency mesh in the actual experimental setup). 

Three figures of merit have been optimized: the probability that the electrons enter the holes ($\mathcal{T}_{in}$), the probability that they punch through them ($\mathcal{T}_{through}$), and the intrinsic energy resolution, $Q$. For completeness, and for a better evaluation of the situation, we consider 5, 10 and 20~mm thickness. The purpose is to illustrate how, in simulation, even such unusually large dimensions still allow electron transmission through the channels, in principle, if a good enough field focusing is guaranteed at the hole entrance (e.g., Fig. \ref{FirstSim}-bottom). The above magnitudes can be defined in simulation as:
\begin{eqnarray}
& \mathcal{T}_{in}      &=  \frac{n_{in}}{n_{o}}   \\
& \mathcal{T}_{through} &=  \frac{n_{out}}{n_{in}} \\
& Q &= \left(\frac{\sigma_{m_\gamma}}{m_\gamma}\right)^2
\end{eqnarray}
where $n_o$ is the number of electrons shot from the cathode, $n_{in}$ the number of them entering the structure (i.e., counted at the entrance plane) and $n_{out}$ is the number of those reaching the anode. The $Q$ factor represents the relative spread (squared) of the scintillation process, characterized by an optical gain $m_\gamma$. 

\begin{figure}[h!!!]
\begin{center}
\includegraphics[width=9.0cm]{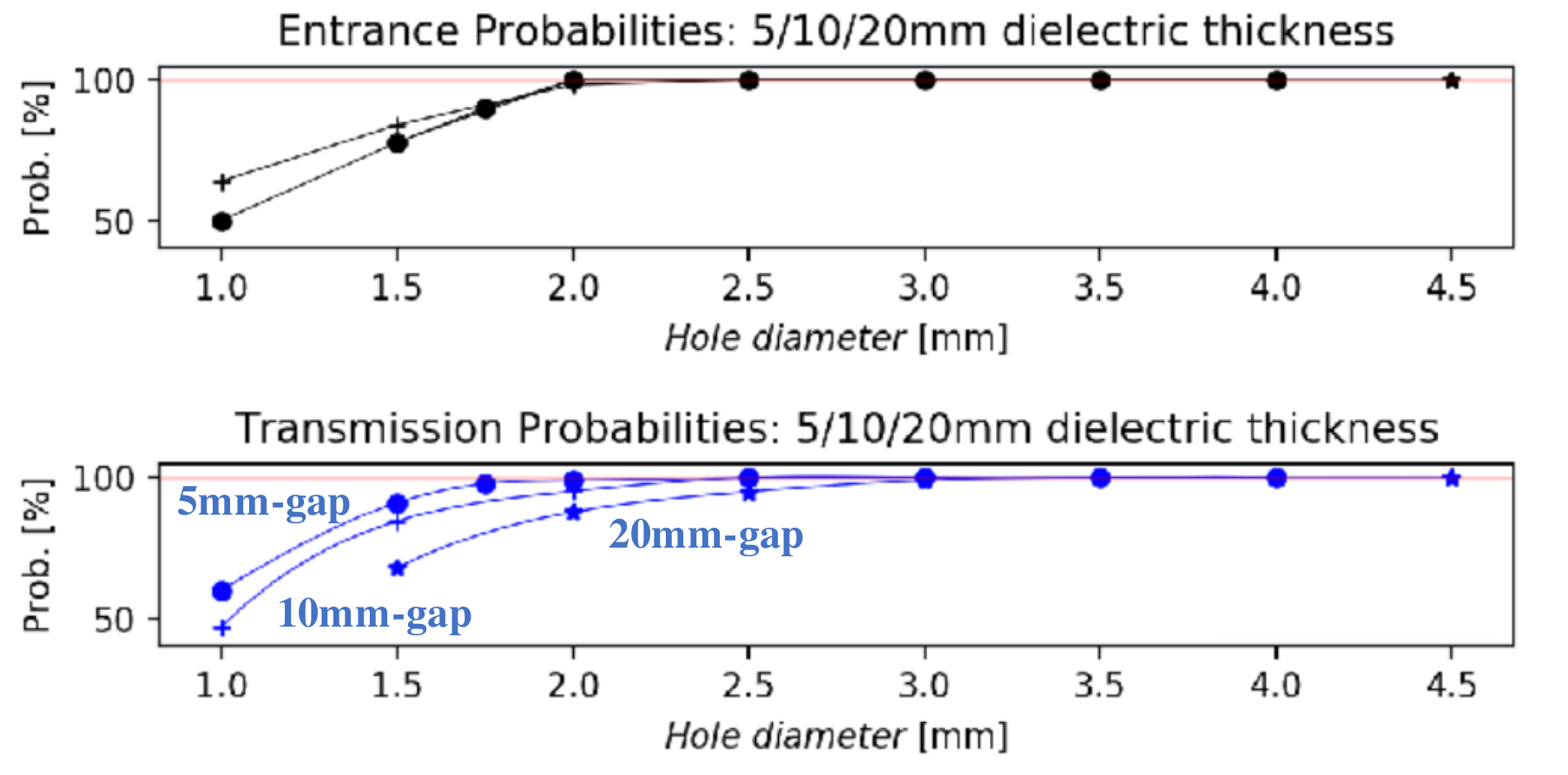}
\includegraphics[width=9.0cm]{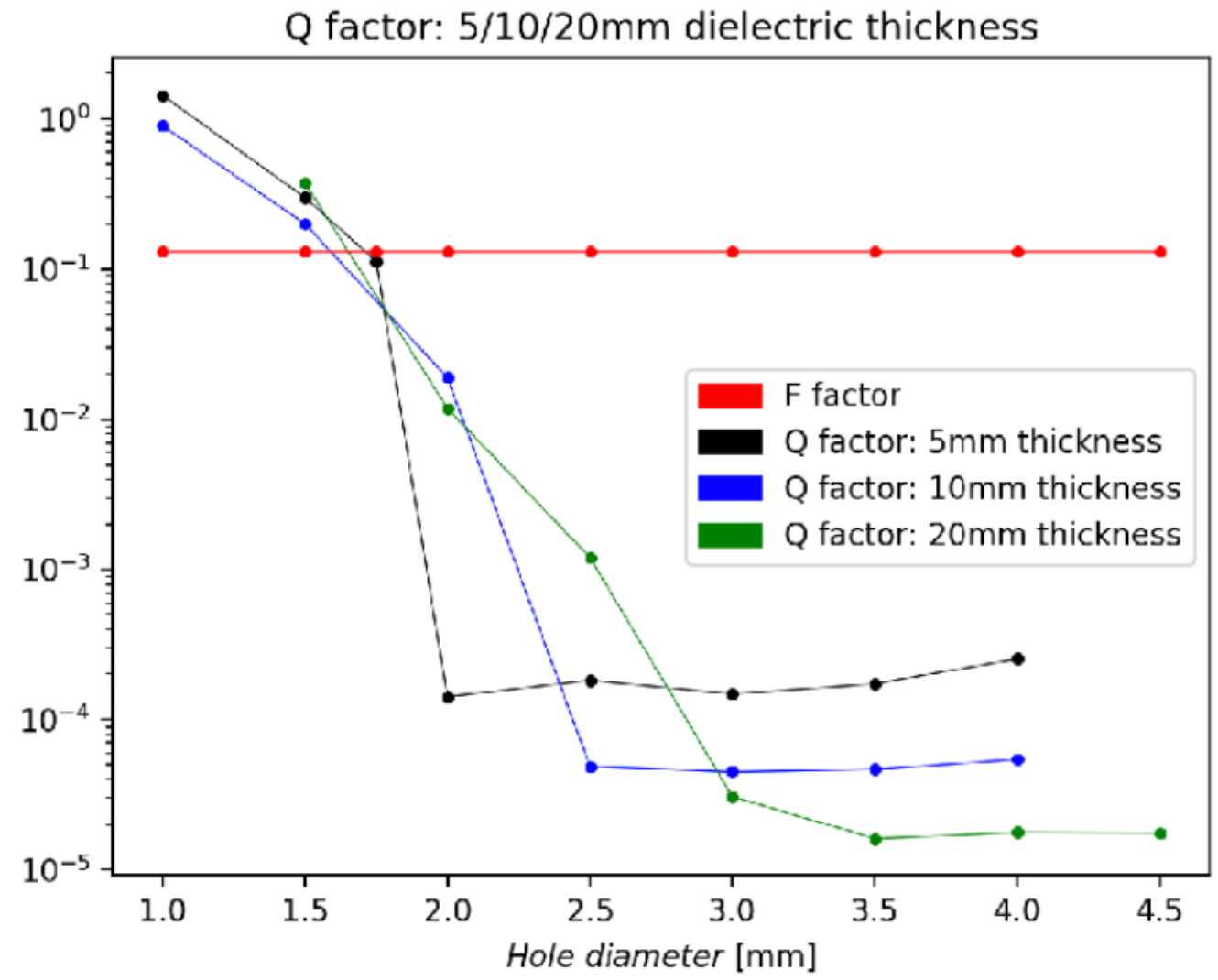}
\end{center}
\caption{\label{Sim_main_par} Top: entrance probability ($\mathcal{T}_{in}$) for different FAT-GEM thicknesses. Middle: transmission probability ($\mathcal{T}_{through}$). Bottom: $Q$-factor (`intrinsic energy resolution') and comparison with the Fano factor (horizontal line). Results obtained with Garfield++.}
\end{figure}

The energy resolution, after including the intrinsic fluctuations in the ionization process and the finite photon statistics in the optical sensor, reads (\cite{Henr}, for instance):
\begin{equation}
\mathcal{R} = 2.355\sqrt{ F + Q + \frac{1}{N_{pe}}\left(1 + \frac{\sigma_{_G}^2}{G^2}\right)}\cdot\frac{1}{\sqrt{N_e}} \label{FormRes}
\end{equation}
where $F$ is the Fano factor, $N_{e}$ the number of ionization electrons, $N_{pe}$ the number of photoelectrons detected per ionization electron and $\frac{\sigma_{_G}}{G}$ represents the relative spread of the multiplication process in the optical sensor (the photomultiplier, in this case). Given that $Q \ll F$ under uniform fields, studying its value represents the most natural way to understand the limitations of a hole structure compared to meshes, that stem from the different trajectories within the hole, and from losses at either the hole entrance or to the hole walls.

The transmission probabilities are compiled in Fig. \ref{Sim_main_par}-top and the $Q$ factor is given in Fig. \ref{Sim_main_par}-bottom. Clearly, above or around 2~mm-diameter holes, simulations predict a near-intrinsic energy resolution in xenon at 10~bar, so we concentrated on this hole size in our experimental studies.

\begin{figure}[h!!!]
\begin{center}
\includegraphics[width=15cm]{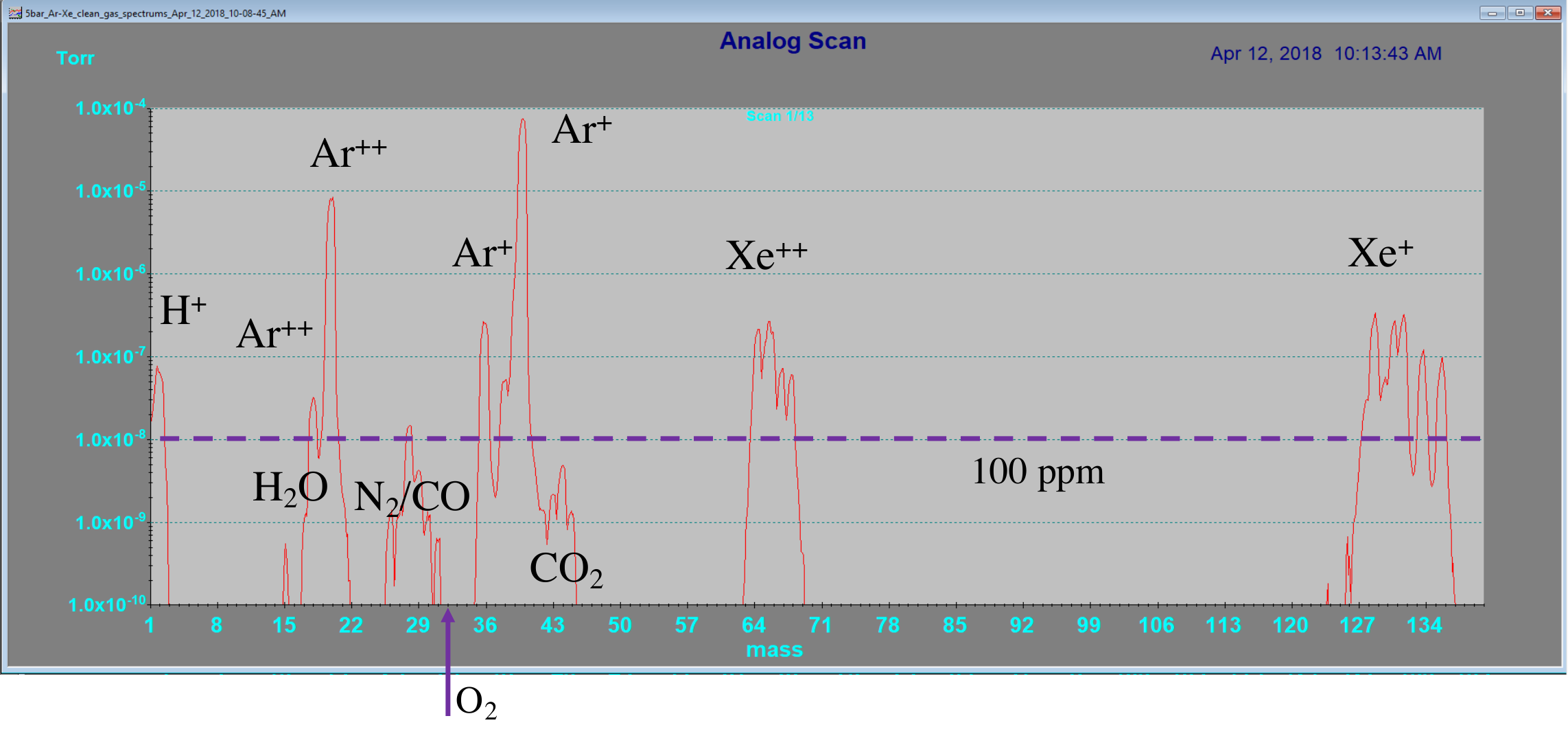}
\end{center}
\caption{\label{RGA} Plot illustrating a gas purity measurement in our experimental setup, done at 5~bar of Ar/Xe (97/3) through an uncalibrated leak in valve.}
\end{figure}

\section{Experimental setup, data analysis and absolute calibration of photon yield}

We commissioned in our lab a new VCR/ConFlat-based gas system equipped with precision mass-flow regulators, the gas being purified through a membrane compressor (all-metal) and a SAES cold getter (MC190). A central vessel (a multi-port CF-100 tube) houses the experimental setup and the gas quality is monitored through a leak valve and a residual gas analyzer (RGA model SRS-200). The system was thoroughly leak-tested with He down to the RGA sensitivity. Once commissioned, the contamination observed in the main gas volume was generally sub-ppm level for O$_2$ and $\lesssim 100$~ppm for N$_2$ and H$_2$O, with a sensitivity limited by the vacuum level achieved in the RGA region (Fig. \ref{RGA}). The measurements presented here were taken along several experimental campaigns, after which the xenon was cryo-recovered.

\begin{figure}[h!!!]
\begin{center}
\includegraphics[width=10cm]{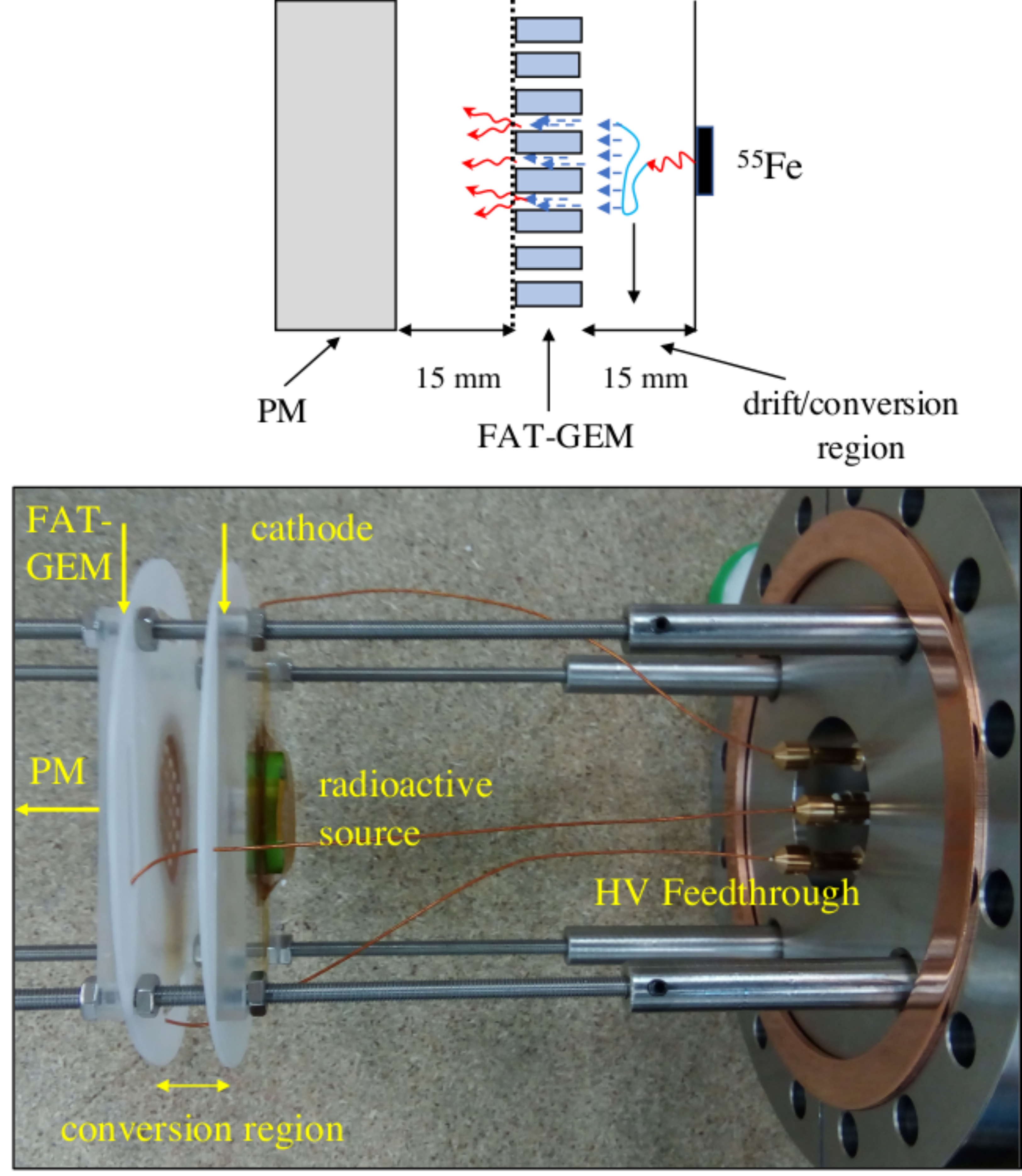}
\end{center}
\caption{\label{setup} Top: sketch of the experimental setup used in present measurements. Bottom: a photograph.}
\end{figure}

\begin{figure}[h!!!]
\begin{center}
\includegraphics[width=14cm]{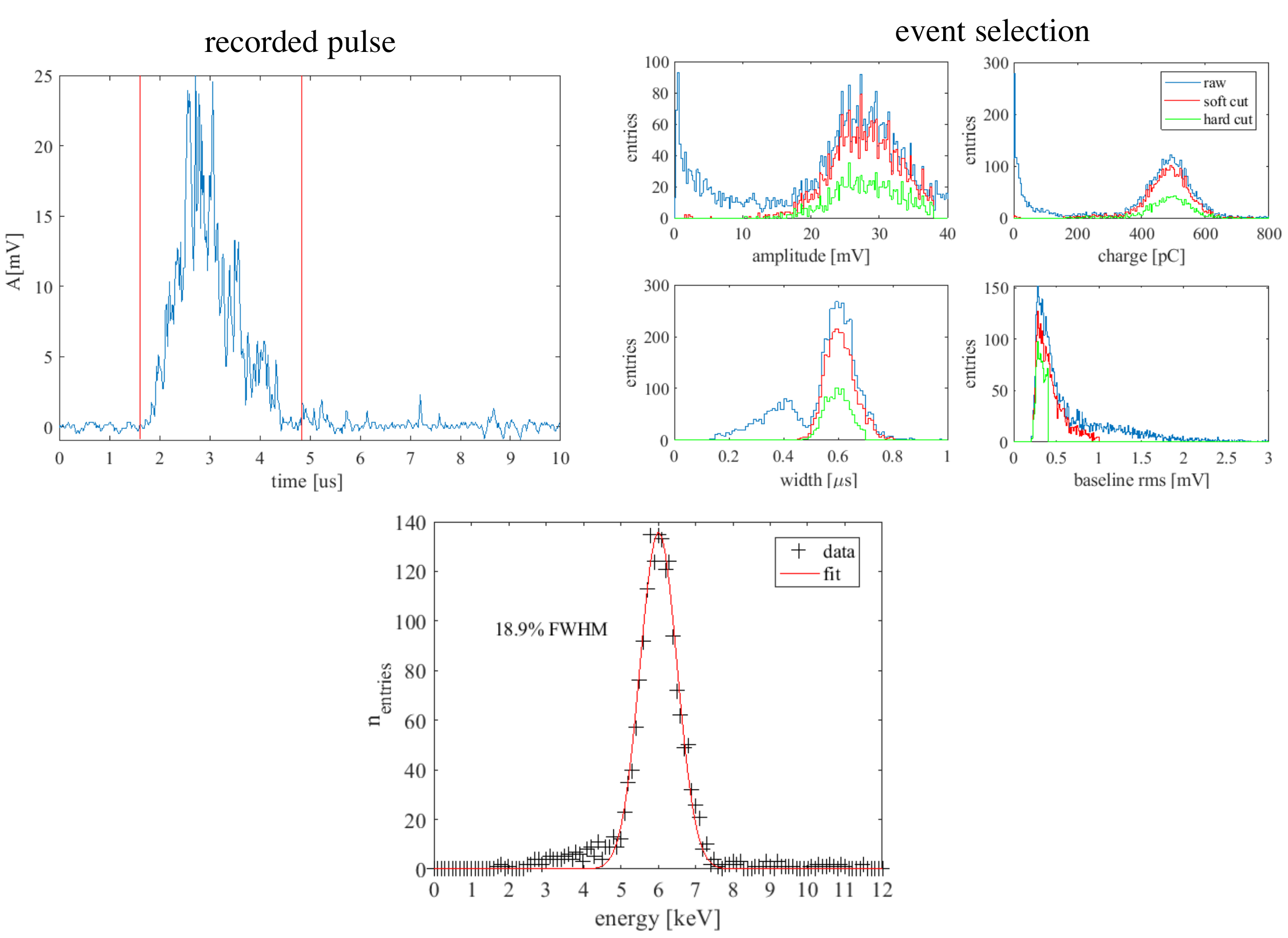}
\end{center}
\caption{\label{analysis} Top-left: a typical $^{55}$Fe x-ray pulse, with a Gaussian shape dominated by diffusion. Top-right: main variables used in the analysis. The low-charge tail can be removed by width and baseline selection. It stems presumably from parasitic scintillation somewhere in the experimental setup. Bottom: an energy spectrum.}
\end{figure}

We resorted to several setups with one and four VUV photo-multipliers (PM-7378, from Hamamatsu), two parallel meshes, FAT-GEMs and different radioactive sources ($^{133}$Ba, $^{55}$Fe, and LEDs), in order to develop a detailed understanding of the system. The final layout chosen for the measurements can be seen in Fig.\ref{setup}-top, consisting of a single PM, centered with respect to a $^{55}$Fe source (activity at the drift/conversion region estimated in around 300-600~Hz). The initial use of polyethylene covers (visible in Fig.\ref{setup}-bottom) was aimed at reducing corona effect from the connection points, but resulted in strong charging-up and transient behaviour of the energy spectra, so they were removed in the final configuration. Data acquisition was implemented using a 1~GHz Tektronix oscilloscope, using its internal 20~MHz filter. A clean sample of events was obtained on the basis of cuts performed on the variables obtained from pulse shape analysis, as sketched on Fig.\ref{analysis}-top. Four main variables were used during the analysis: amplitude, charge, width and a metric to characterize the baseline fluctuations. For better event selection and aiming at the best energy resolution, cuts were applied in the event width and baseline fluctuations, although the impact on the final energy resolution was small.

\begin{figure}[h!!!]
\begin{center}
\includegraphics[width=15cm]{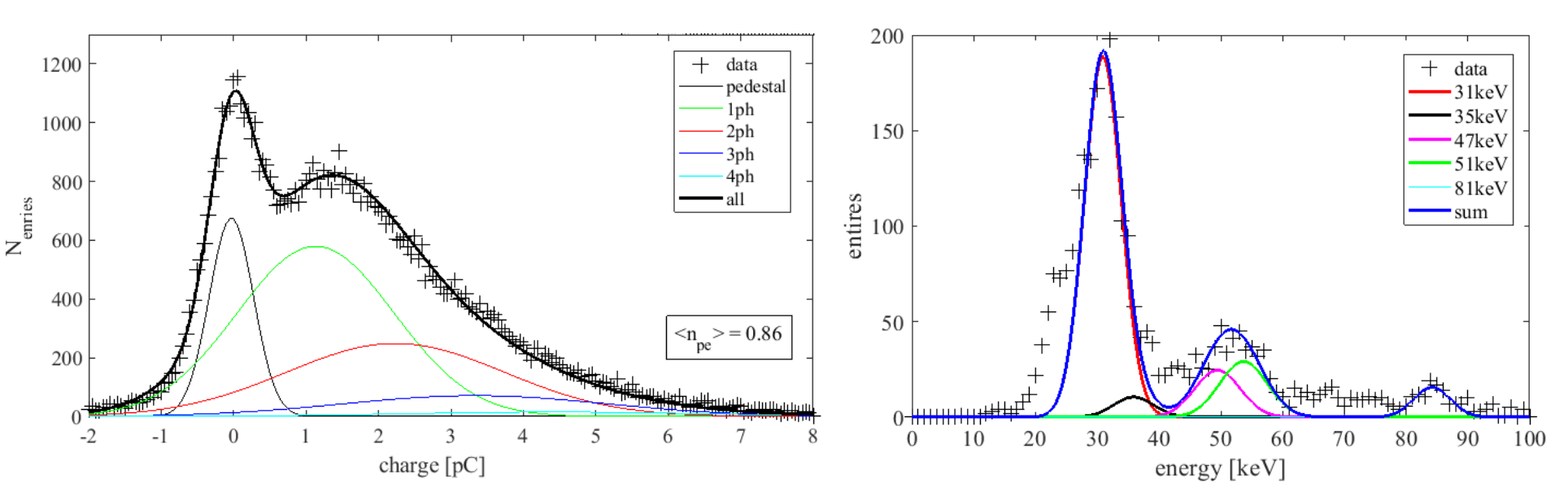}
\end{center}
\caption{\label{Cal} Left: LED calibration (measurements done for different number of photoelectrons, 0.86 in the plot, yield fit parameters compatible within 10\%). Right: measurement of x-rays from a $^{133}$Ba source performed with two meshes and four PMs. The position of the peak was used to estimate the charge-to-photon conversion based on previous EL measurements \cite{Monteiro}.}
\end{figure}

Two types of calibration data were taken, one with two meshes at 5~mm distance and a $^{133}$Ba source, and one with a LED (Fig.\ref{Cal}). After assuming an effective quantum efficiency of QE=0.05 to the second continuum of xenon (172~nm), and correcting for mesh transparency and solid angle, comparison with previous EL-measurements and simulations \cite{Henr} provided a calibration of about 1~pC/photon, in agreement with the one obtained with the LED, within 20\%. This agreement further supports, as hinted by the RGA measurements, that the influence of impurities in the scintillation is of negligible importance in our system.

\section{Results}

\begin{figure}[h!!!]
\begin{center}
\includegraphics[width=14.5cm]{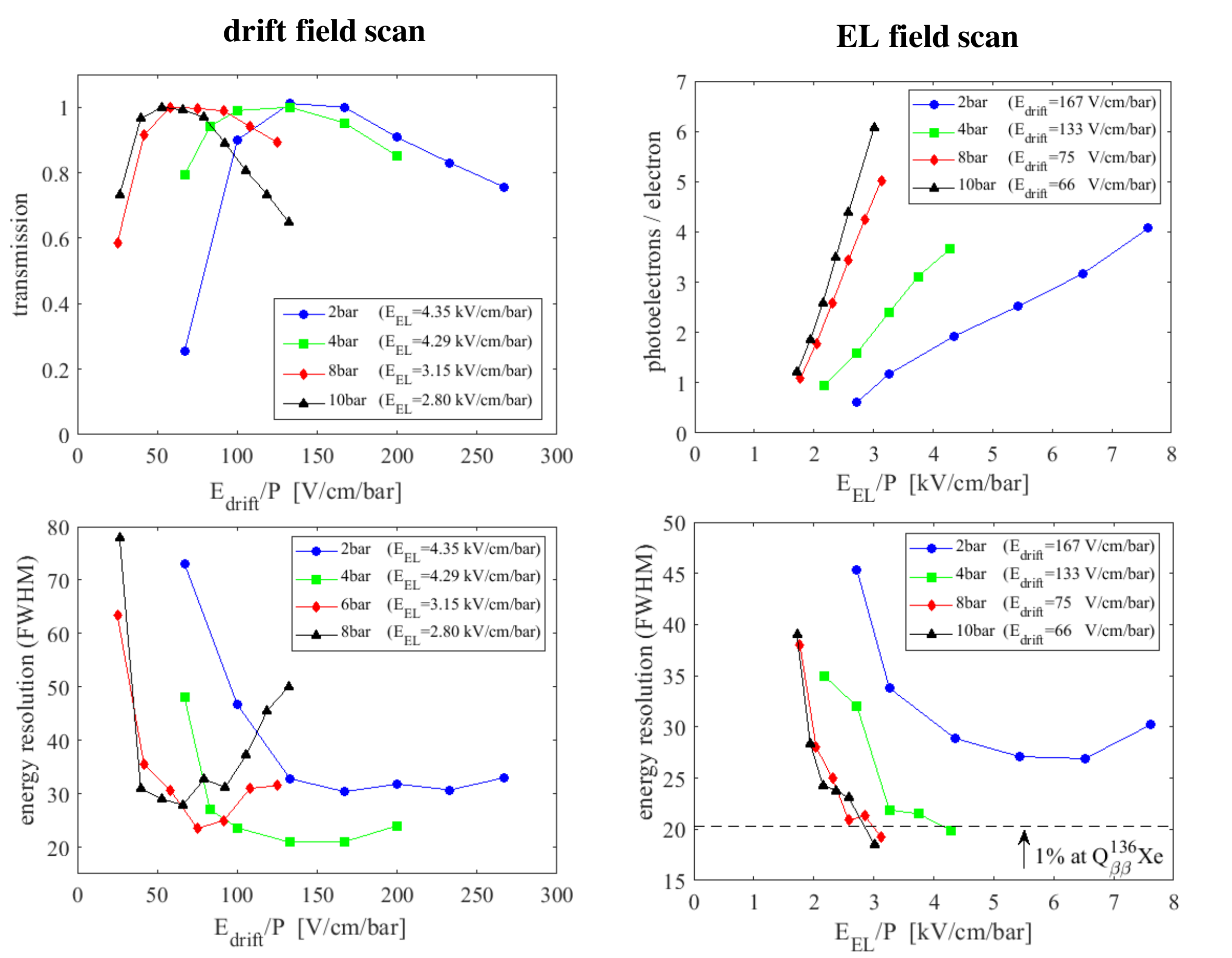}
\end{center}
\caption{\label{EL_vs_field} Top-left: charge collection as a function of the reduced drift field, normalized to the maximum value. Bottom-left: energy resolution in the same conditions as before. Top-right: number of photoelectrons per electron ($N_{pe}$) as a function of the reduced electroluminescence field. Bottom-right: energy resolution in the same conditions as before.}
\end{figure}

Data were taken at 2, 4, 8 and 10 bar. Fig. \ref{EL_vs_field}-top/left shows a characteristic `electron transmission' curve (arbitrarily normalized to one at the maximum), namely, the position of the x-ray peak as a function of the drift field. The drop observed at around 75~V/cm/bar (10~bar) is in approximate agreement with the one observed in simulation, and can be attributed to the loss of field focusing at high drift fields. The drop at low fields, on the other hand, is difficult to explain given that no charge recombination has been observed earlier in these conditions (e.g., \cite{Balan}). The effect is weaker at high pressure, that is hardly compatible with attachment; finally, there is no particular feature in the diffusion coefficients that could explain it \cite{Austin}. We suspect charging-up of the region outside the 55~mm-diameter electrodes, similar to the effect seen earlier with the polyethylene covers, but milder. This can create a field of opposite sign, reducing the total field with respect to the estimate made from the voltage difference between cathode and GEM entrance, and explains why charge losses are stronger for lower voltage difference (i.e., low pressure). The energy resolution deteriorates accordingly at both low and high fields, as indicated in Fig.\ref{EL_vs_field}-bottom/left.

\begin{figure}[h!!!]
\begin{center}
\includegraphics[width=8cm]{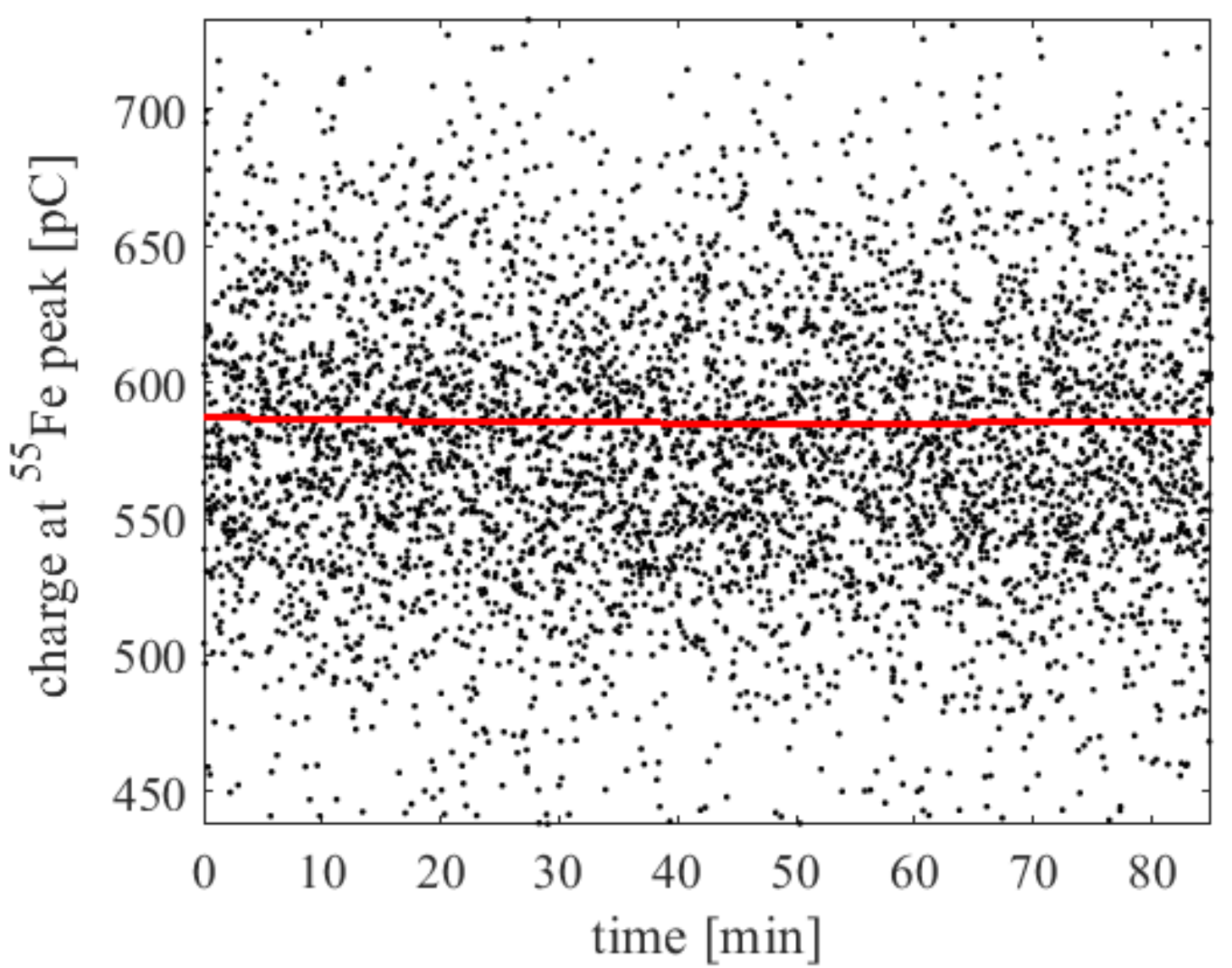}
\end{center}
\caption{\label{transient} Charge at the $^{55}$Fe peak as a function of the measuring time.}
\end{figure}

Following the above observation, during the EL-scan we kept $E_d$ constant, instead of keeping constant the ratio $E_d/E_{el}$, as customary when charge collection is dominated by losses at the holes entrance \cite{PacoGreat}.  The EL-scan is shown in Fig. \ref{EL_vs_field}-top/right after calibration, in photoelectron units. The energy resolution is given in Fig. \ref{EL_vs_field}-bottom/right, indicating the value that extrapolates to 1\% level at the $Q_{\beta\beta}$ of xenon (2.45~MeV), the target resolution of NEXT. During the measurements, no transient effect was observed (Fig. \ref{transient}).

\begin{figure}[h!!!]
\begin{center}
\includegraphics[width=13.7cm]{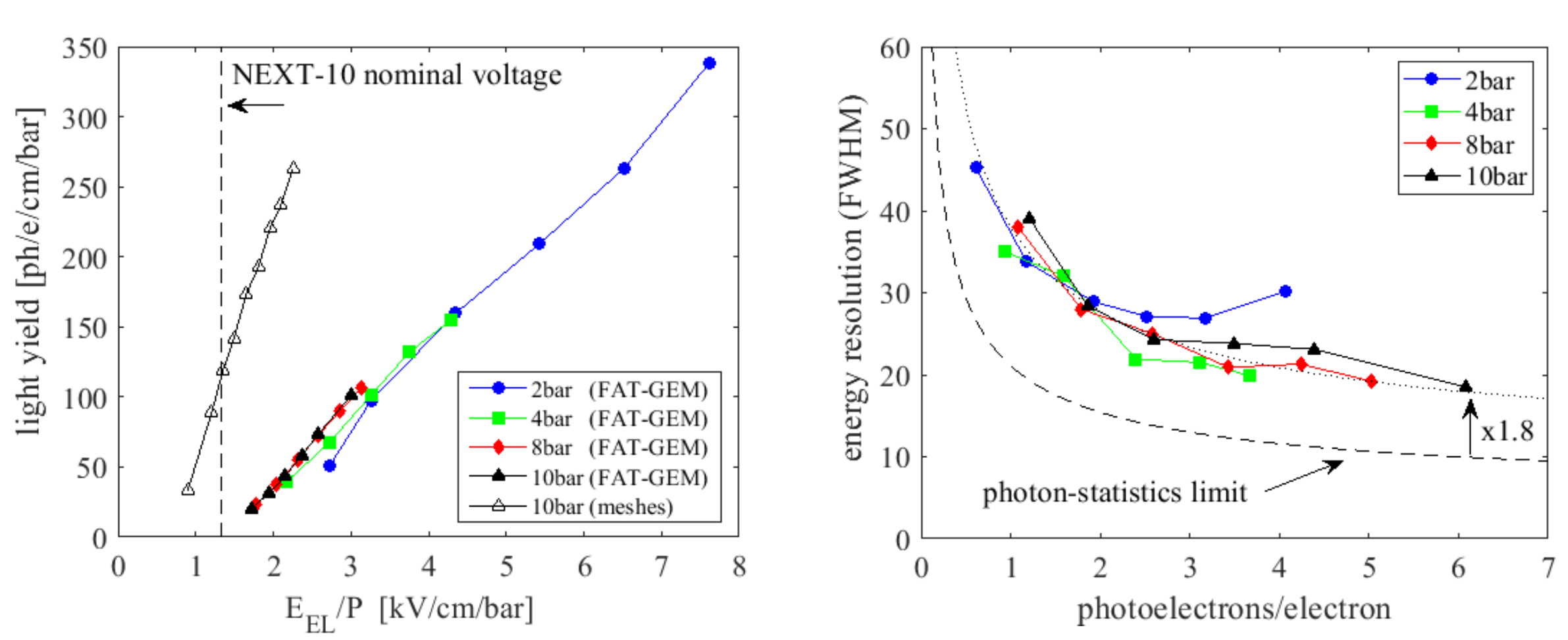}
\end{center}
\caption{\label{EL_comp} Left: reduced scintillation yield for meshes (open symbols) from \cite{Monteiro} and from present structure, as a function of the EL-field and at various pressures. Right: energy resolution as a function of the number of photoelectrons per electron ($N_{pe}$). The photon statistics limit (dashed lines) results from the evaluation of eq. \ref{FormRes} with $Q\ll{F}$, and the dotted line is the same curve after multiplying by a factor $\times 1.8$.}
\end{figure}

Fig. \ref{EL_comp} shows a comprehensive summary of the main results obtained in this work. To the left, we compare the reduced scintillation yield $Y$ (in ph/e/cm/bar) with the one expected from meshes, after solid angle correction:
\beq
Y = N_{pe}/(\mathcal{T}\cdot{\Omega}\cdot{t})
\eeq
where $\mathcal{T}$ is the mesh transparency, $\Omega$ the solid angle subtended by the PM at a point placed at the FAT-GEM mid-plane, and $t$ is the FAT-GEM thickness (4.6~mm). Clearly, a large reduction is observed, due to the collimation effect of the acrylic material, that amounts to a factor $\times 1/5$ at the highest fields. The threshold for electroluminescence appears at around 1.2~kV/cm/bar, that is hence the minimum field required to achieve full electron transmission at typical drift fields (50~V/cm/bar at 10~bar). Despite this reduction, the optical gain is at the level of the one observed currently in NEXT, given the much higher operating voltage achieved for FAT-GEMs, and equals $m_\gamma = 500 $~ph/e in present experimental conditions. Moreover, the maximum stable voltage across the structure is limited at 10~bar by the power supplies and feedthroughs, to about 13~kV, and not by the structure.

In order to investigate the present limitations on the energy resolution, Fig. \ref{EL_comp}-right shows its behaviour as a function of the number of photoelectrons per electron ($N_{pe}$). It is interesting to note the deviation from the $1/\sqrt{N_{pe}}$ scaling for the lowest pressure and highest field (corresponding to $E_{EL}/P=7.5$ kV/cm/bar), that hints to the onset of avalanche multiplication. The rest of points show an approximate scaling, almost a factor of $\times 2$ in excess of the expectation from the photon-statistics term in eq. \ref{FormRes}. This result encourages us to dedicate additional efforts to improve light collection, minimize possible fringe fields in the drift region, as well as exploring FAT-GEMs with different hole sizes to exclude any effect related to losses.

\section{Conclusions}
We introduced a new electroluminescence structure based on acrylic and relying on MPGD fabrication techniques (dubbed FAT-GEM), capable of reliably operating in xenon at 10~bar, with an optical gain of 500~ph/e, and an energy resolution of 18.9\% for 5.9~keV x-rays. Operation was stable at 3~kV/cm/bar, limited by our experimental setup. The structure is extremely robust: at the lowest pressure, where its breakdown field could be reached, it was exposed repeatedly to sparks without any indication of damage. It withstands cryogenic cycles (10 at least) of LN, without any indication of electrode detachment or fracture.

Assuming energy scaling, these first results are already at the level of the best performance obtained with MPGDs in high pressure xenon to date: Micromegas in Xe-TMA (9.6\%(22.1keV), P=10bar \cite{Diana}) and teflon-based hole structures (7.3\%(28.8keV), P=4bar, \cite{AXEL}).
\ack
DGD acknowledges the Ramon y Cajal program, contract RYC-2015-18820. This work is supported by the RD51 `common projects' initiative. We thank our NEXT colleagues for interesting discussions, and in particular the technical assistance by S. C\'arcel and A. Mart\'inez.

\end{document}